\RequirePackage{fix-cm}
\documentclass[onecolumn,epjc3]{svjour3}  
\smartqed  %
\RequirePackage{graphicx}

\usepackage{amsmath}
\usepackage{amssymb}
\usepackage{epstopdf}
\usepackage{color}
\usepackage{mathptmx}
\DeclareGraphicsExtensions{.eps,.pdf,.png,.gif,.jpg,.mht}

\journalname{Jinst}

\date{\today}

\pdfoutput=1 

\usepackage{graphicx}
\graphicspath{{final-figures/}}
\DeclareGraphicsExtensions{.pdf,.png,.gif,.jpg,.mht}

\usepackage{multirow}
\usepackage{amssymb}
\usepackage{verbatim}
\usepackage{hyperref}

\usepackage{topcapt}

\begin{document}


\title{\boldmath Test of UFSD Silicon Detectors for the TOTEM Upgrade Project}

\author{
R.~Arcidiaconoi\thanksref{a0} \and
 M.~Berretti\thanksref{a1} \and
 E.~Bossini\thanksref{5,4} \and 
 M.~Bozzo\thanksref{3} \and
 N.~Cartiglia\thanksref{i}\and
 M.~Ferrero\thanksref{3c,i}\and
 V.~Georgiev\thanksref{2} \and
 T.~Isidori\thanksref{4} \and
 R.~Linhart\thanksref{2} \and
 N.~Minafra\thanksref{5c} \and
 M.~M.~Obertino\thanksref{3c} \and
 V.~Sola\thanksref{i}\and
 N.~Turini\thanksref{6c,7c}
}

\institute{
Universit\`a degli Studi del Piemonte Orientale, Largo Donegani 1  - 28100 Novara, Italy. \label{a0} \and 
University of Helsinki and Helsinki Institute of Physics, P.O. Box 64, FI-00014, Helsinki, Finland. \label{a1} \and
Universit\`a degli studi di Pisa and INFN di Pisa, Largo B. Pontecorvo 3, 56127 Pisa, Italy. \label{4} \and
Centro Studio e Ricerche Enrico Fermi Piazza del Viminale 1 - 00184 Roma (Italy). \label{5}\and
INFN Sezione di Genova, Via Dodecaneso 33, 16136 Genova, Italy. \label{3} \and
Sezione  INFN di Torino, Via P.Giuria, 1 - 10125 Torino, Italy.\label{i} \and
Universit\`a degli Studi di Torino, Via P.Giuria, 1 - 10125 Torino, Italy\label{3c} \and
University of West Bohemia, Pilsen, Univerzitni 8, PILSEN 30614 (Czech Republic). \label{2} \and
University of Kansas, 1246 West Campus Road, Lawrence - KS 66045, USA. \label{5c} \and
Universit\`{a} degli Studi di Siena and Gruppo Collegato INFN-Siena, Via Roma 56, 53100 Siena, Italy. \label{6c}\and
CERN, Geneva, Switzerland. \label{7c}
}

%
%

%
\thankstext{e1}{Corresponding author's e-mail: mirko.berretti@cern.ch}
\maketitle
%


%

\abstract{This paper describes the performance of a prototype timing detector,  based on 50\,$\mu$m  thick Ultra Fast Silicon Detector, as measured in a beam test using a 180\,GeV/c momentum pion beam.
The dependence of the time precision on the pixel capacitance and the bias voltage is investigated here. 
A timing precision from 30\,ps to 100\,ps, depending on the pixel capacitance, has been measured at a bias voltage of 180 V. 
Timing precision has also been measured as a function of the bias voltage.
}







\section{Timing Detector for the TOTEM  Proton Time of Flight  Measurement at the LHC}
\label{sec:det}

The TOTEM experiment will install new timing detectors to measure the time of flight (TOF) of protons produced in central diffractive (CD) collisions at the LHC~\cite{Albrow:1753795}.\\

The CD interactions measured by TOTEM at $\sqrt{s}=13$\,TeV are characterized by having two high energy protons (with momentum  greater than 5 TeV) scattered at less than 100\,$\mu$rad from the beam axis. 
In the presence of pile-up\footnote{Probability that more than one interaction is produced during the same bunch crossing.} events  the reconstruction of the protons interaction vertex position allows to associate the physics objects reconstructed by the CMS experiment with the particles generated from that vertex.
The TOF detectors installed in the TOTEM  Roman Pots (RPs)\footnote{Special movable insertion in the LHC vacuum beam pipe  that allow to move a detector edge very close to the circulating beam.} will measure with high precision  the arrival time of the CD protons on each side of the interaction point.
They will operate in  the LHC with a scenario of moderate pile-up ($\mu\sim$1) and a time precision of at least 50\,ps per arm is required to efficiently identify the event vertex~\cite{CERN-LHCC-2014-024}.
Since the difference of the arrival times is directly proportional to the longitudinal position of the interaction vertex ($z_{VTX} = c \Delta t/2$), a precision of 50\,ps will allow to know the longitudinal interaction vertex position to less than 1\,cm.

The timing detector will be installed in four vertical RPs located at 210\,m from the interaction point 5 (IP5) of the LHC.
The detector comprises four identical stations, each consisting of four hybrid boards\footnote{The particle sensor and the amplification electronic are mounted on the same PCB.} equipped either with  an ultra fast silicon detector (UFSD)~\cite{DallaBetta2015154}, \cite{Sadrozinski20147}, \cite{Sadrozinski2013226},\cite{Cartiglia2015141} or with a single crystal chemical vapor deposition (scCVD) diamond sensor~\cite{timing-nov-15}, \cite{Berretti:2016sfj}. 
The board contains 12 independent amplifiers, each bonded to a single pad (pixel) of the sensors. 
The typical time precision of one  plane equipped with scCVD  is in the range of 50 − 100\,ps,  while it is in the 30 − 100\,ps range for one equipped with an  UFSD sensor.
Combining TOF measurements from 4 detector planes will provide an ultimate time precision better than $\sim$50\,ps, which translates in a precision on the longitudinal position of the interaction vertex $\sigma_z\,<$1\,cm. 

\section{Ultra Fast Silicon Detector}
\label{sec:ufsd}
Ultra Fast Silicon Detectors, a new concept in silicon detector design, associate the best characteristics of standard silicon sensors with the main feature of Avalanche Photo Diodes (APD).

\begin{figure}[!h]
\begin{center}
\includegraphics[width=0.7\textwidth]{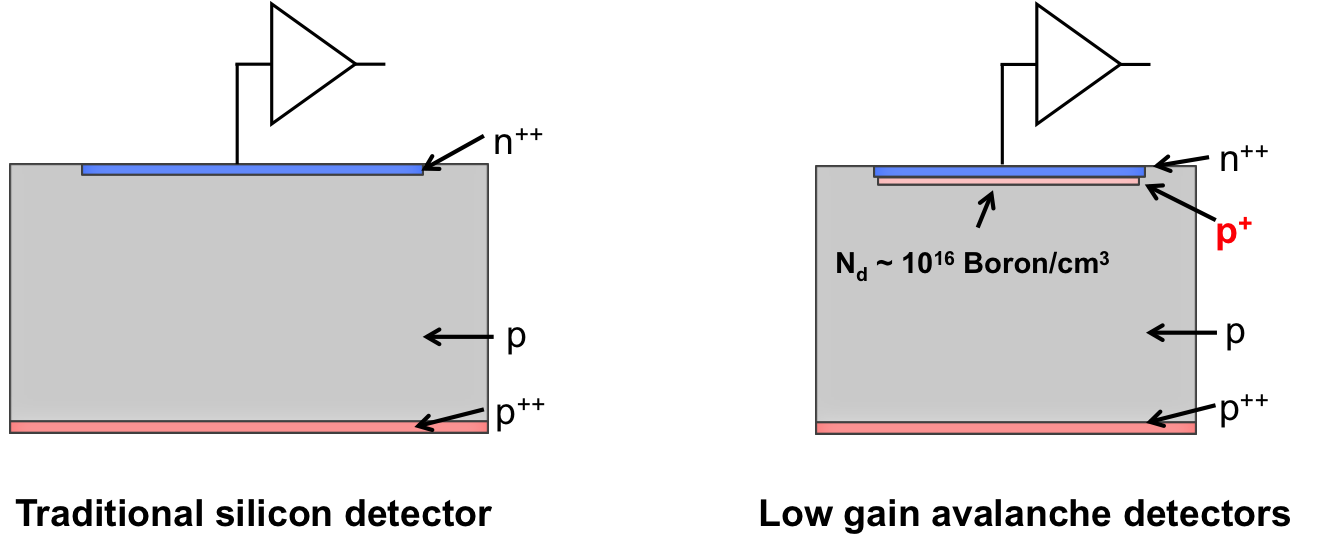}
\caption{Comparison of the structures of a  silicon diode (left) and a Low-Gain Avalanche Diode (right). The additional $p^+$ layer near the $n^{++}$ electrode creates, when depleted, a large electric field that generates charge multiplications.}
\label{fig:LGAD}
\end{center}
\end{figure}
 
UFSD are thin (typically $50 \mu$m thick) silicon Low Gain Avalanche Diodes (LGAD)~\cite{FernandezMartinez201198}, \cite{Pellegrini201412}, that produce  large signals showing hence a large $dV/dt$, a characteristic necessary to measure time accurately.

Charge multiplication in silicon sensors happens when the charge carriers drift in electric fields of the order of $E \sim 300$\,kV/cm. 
Under this condition the drifting electrons acquire sufficient kinetic energy to generate additional e/h pairs. 
A field value of 300 kV/cm in a semiconductor can be obtained by implanting an appropriate charge density around $N_D \sim 10^{16}/cm^3$, that will  locally generate the required very high fields. 
Indeed in the LGAD design (Figure~\ref{fig:LGAD}) an additional doping layer is added at the $n-p$ junction which, when fully depleted, generates the high field necessary to achieve charge multiplication.
Gain depends exponentially on the value of the electric field E, $ N(l) = N_o e^{\alpha (E)l}$, where $\alpha$ is a strong function of E and $l$ is the mean path length in the high field region. 

First results of time resolution of thin LGADs (UFSD),  and based on beam test measurements, have been published in 2016~\cite{Cartiglia:2016voy}.

\begin{figure}[!h]
\centering
\includegraphics[width=0.65\linewidth,height=0.4\textheight]{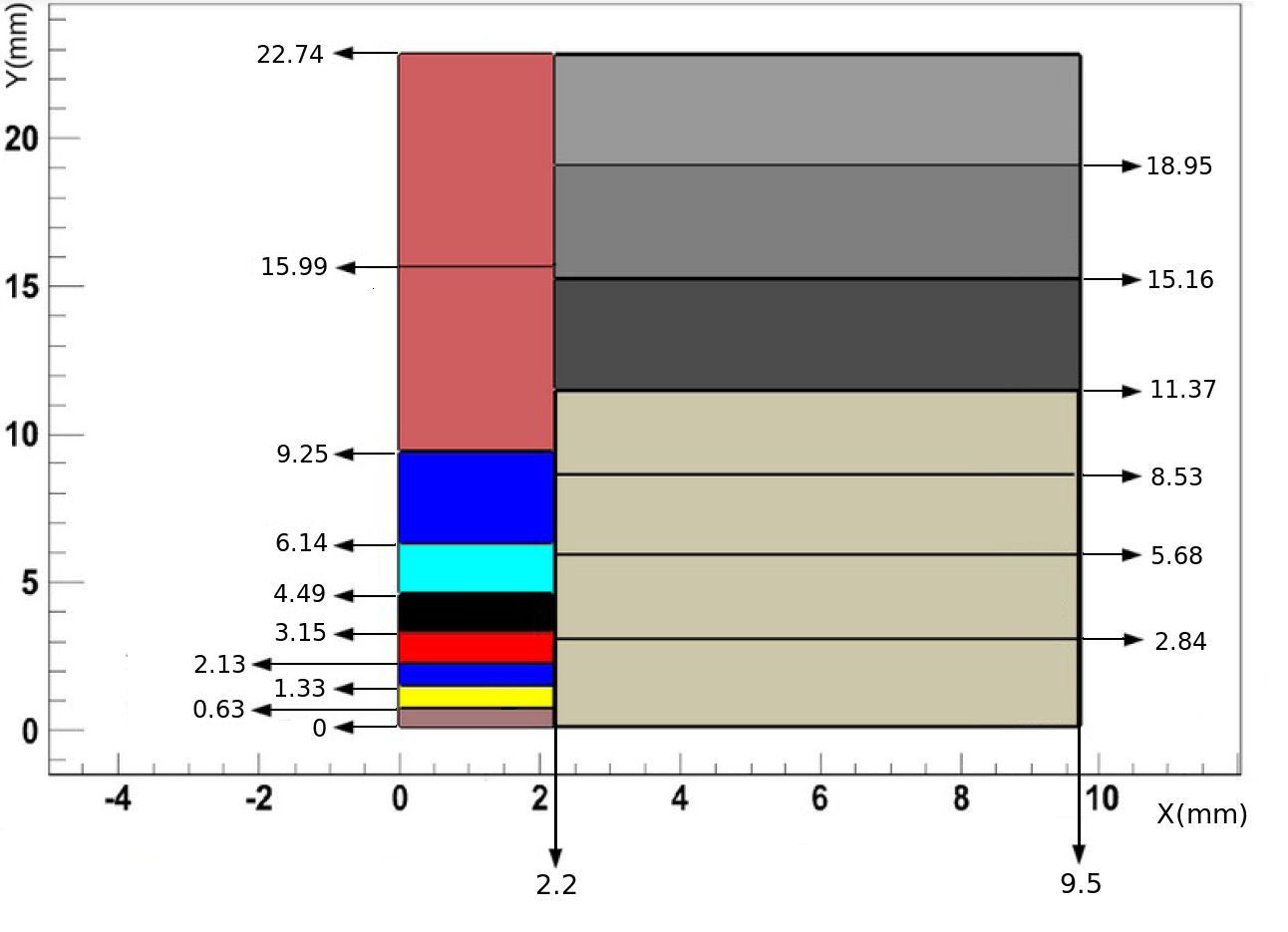}
\caption{Sensor geometry of the TOTEM UFSD prototype made up of 15 pixels of different dimensions.}
\label{geogeo}
\end{figure}


Radiation tolerance studies have shown~\cite{Baldassarri2016}, \cite{1748-0221-10-07-P07006} that LGAD sensors can withstand up to $10^{14} \; n_{eq}/cm^2$ without loss of performance. 

LGAD sensors can be built in many sizes and shapes, ranging from thin strips to large pads. 
The measurements  reported here have been performed on a 2 cm$^2$ 50$\mu$m thick UFSD sensor,  manufactured  by CNM\footnote{\href{http://www.cnm.es} Centro Nacional de Microelectr\'onica, Campus Universidad Aut\'onoma de Barcelona.  08193 Bellaterra (Barcelona), Spain.} with a structure specifically designed for the TOTEM experiment, mounted on a standard TOTEM hybrid board~\cite{timing-nov-15}.

\section{Description of the UFSD-based  Timing Board}
\label{sec:descr}
 
The UFSD sensor used for the prototype timing plane has 15 pixels  with the pixel layout shown in Figure~\ref{geogeo}.

\begin{table}[!h]
\centering
\caption{Characteristics of the 50\,$\mu$m UFSD pixels used in the tests.}
\begin{tabular}{c c c c}
\hline
Pixel N & Surface  & Capacitance & Preamplifier feedback\\
 & [mm$^2$] & [pF] & [ohm]\\
 \hline

1 & 1.8 & 3.1 & 1 k\\
2 & 2.2 & 4.4 & 1 k \\
3 & 3.0 & 6.0 & 1 k \\
4 & 7.0 & 14 & 1 k \\
5 & 14 & 28  & 300  \\

\hline
\end{tabular}

\label{table:dimensions}
\end{table}

Prior to the gluing of the sensor on the hybrid board, each of the 15 pixels  had been tested in the lab to determined its maximum operating voltage.

\begin{figure}[!hbt]
\centering
\includegraphics[width=0.4\linewidth, trim=2cm 8cm 5cm 2cm, clip]{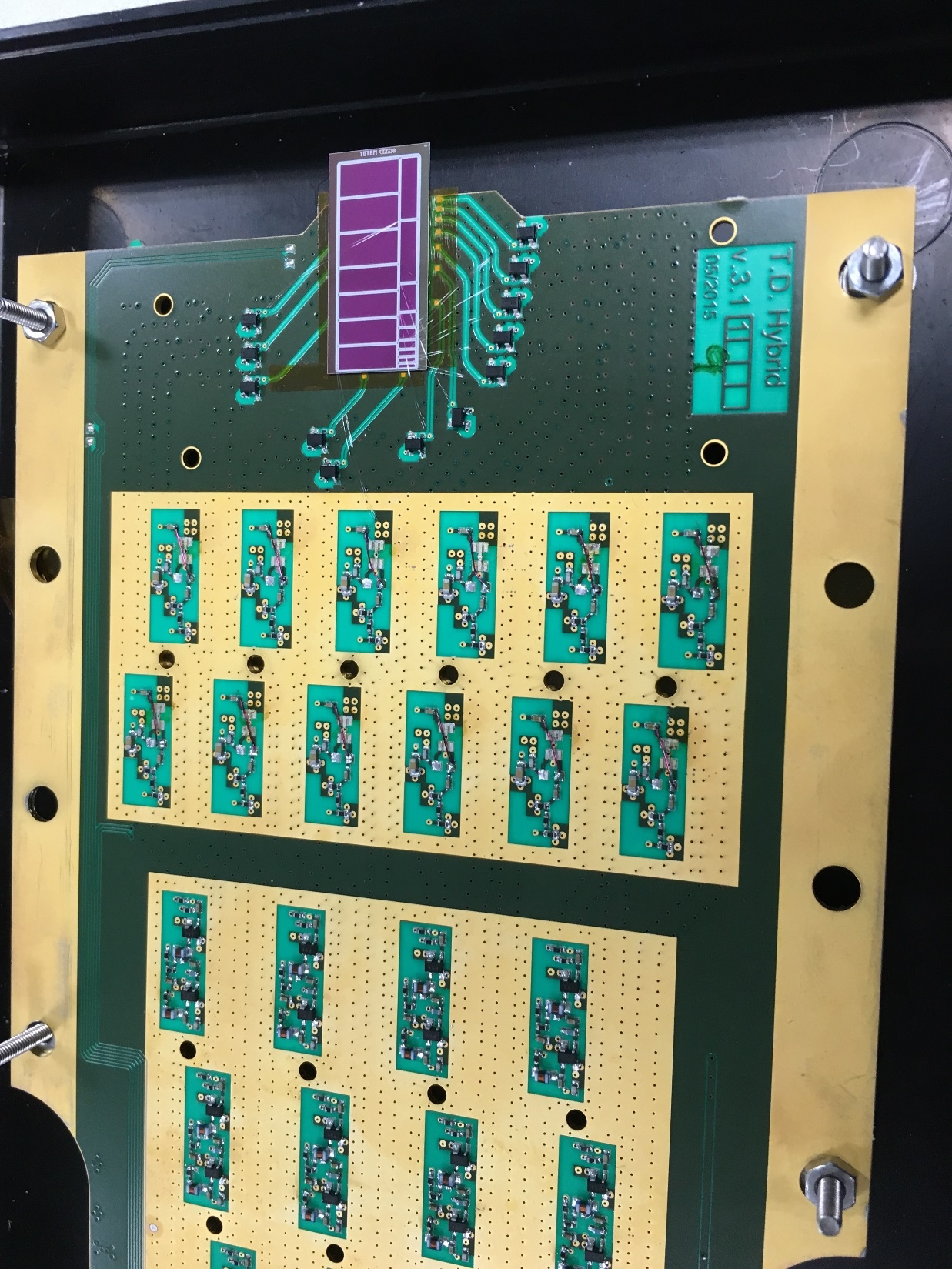} 
\caption{The UFSD sensor mounted  on the TOTEM hybrid board.}
\label{pianlab}
\end{figure}

Only pixels with a breakdown voltage higher than 180\,V and a leakage  current lower than 0.1\,mA, were bonded to the amplification channel by means of standard 25\,$\mu m$ aluminum wires (Figure~\ref{pianlab}). 

\begin{figure}[h]
\centering
\includegraphics[width=0.58\linewidth]{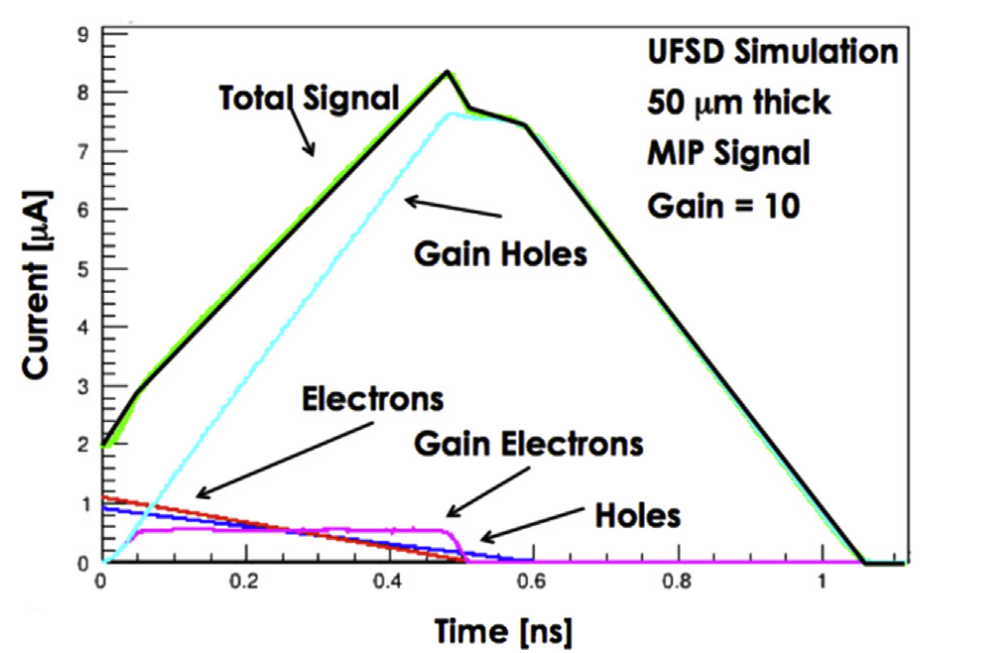}
\caption{Simulations of the pulse shape from a 50$\,\mu$m UFSD  with a gain of 10 (from~\cite{Cartiglia:2015iua}). The plot shows the contribution of each component of the generated charge.}
\label{currr}. 
\end{figure}

The UFSD output pulse shape simulated with the simulation program Weightfield2\footnote{open source code may be found at~\url{http://personalpages.to.infn.it/~cartigli/Weightfield2/Main.html}}, developed particularly for LGAD devices~\cite{Cenna2015149}, assuming a bias voltage of 200\,V and a sensor gain of 10 is shown in Figure~\ref{currr}.

The detector generates a current whose maximum is about 8\, $\mu$A.

Capacitance of the 50\,$\mu$m thick UFSD  pixels scales linearly with their area as $\sim\,$2 pF/mm$^2$: dimensions and relative capacitance for the pixels measured here  are summarized in table~\ref{table:dimensions}.


\section{Front End Electronics}
\label{sec:Feel}
Given the UFSD intrinsic charge amplification one expects the primary charge presented at the input of the amplifier to be 10-100 times larger than the one expected from a diamond sensor.
The TOTEM hybrid, originally designed for scCVD diamonds~\cite{timing-nov-15}, was modified for the UFSD eliminating the second amplification stage, referred elsewhere as ABA. 
The amplification chain has now only 3 active elements (one BFP840ESD and two BFG425W BJT 
transistors).
\begin{figure}[!h]
\centering
\includegraphics[width=0.6\linewidth]{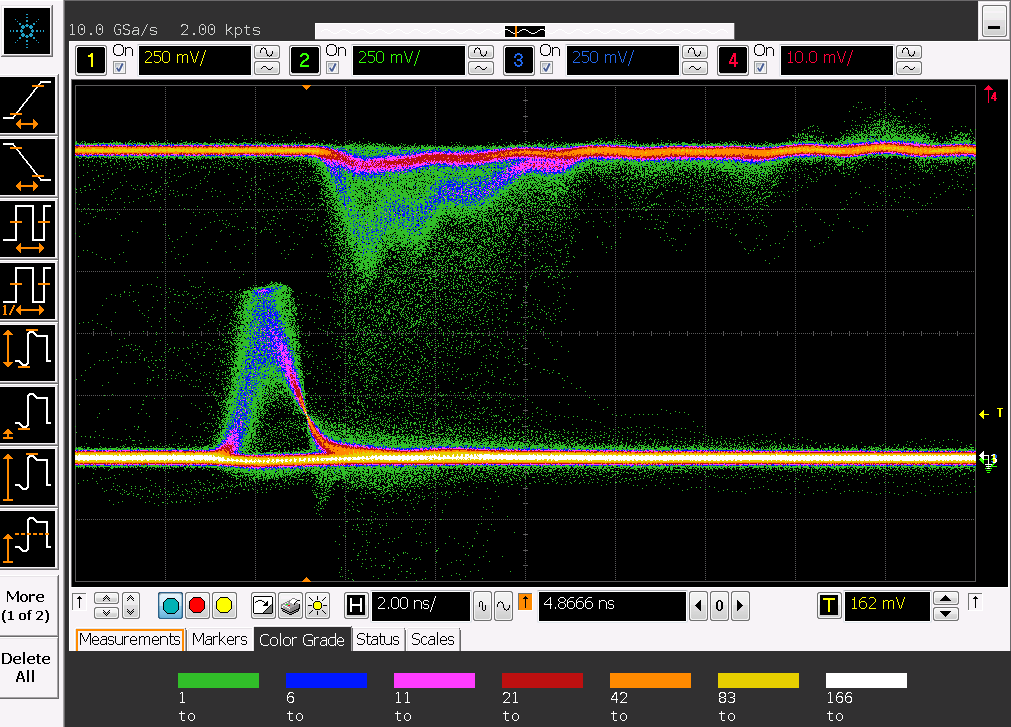}
\caption{Event display of several MCP (top) and UFSD (bottom) signals. The oscilloscope record was triggered by the UFSD signal.}
\label{edisp}
\end{figure}
Moreover, since the UFSD pixels  have a larger capacitance than diamond sensors, in order to maintain a fast rise time the feedback resistor of the preamplification chain has has been  reduced  to to 1k$\Omega$ or 300$\Omega$, accordingly to the capacitance of the pixel (see table~\ref{table:dimensions}).


\section{Test Beam Measurements}
\label{sec:meas}
The time precision of the UFSD sensors has been measured at the H8 beam line of the CERN SPS, a 180 GeV/c  pion beam, by computing the time difference of the signal produced by particles crossing a Micro Channel Plate (MCP) PLANACON$^{TM}$ 85011-501~\footnote{ PLANACON$^{TM}$ Photomultiplier tube assembly 85011-501 from BURLE.}  and one of the UFSD pixels.
\begin{figure}[b]
\centering
\includegraphics[width=0.7\linewidth,height=0.3\textheight]{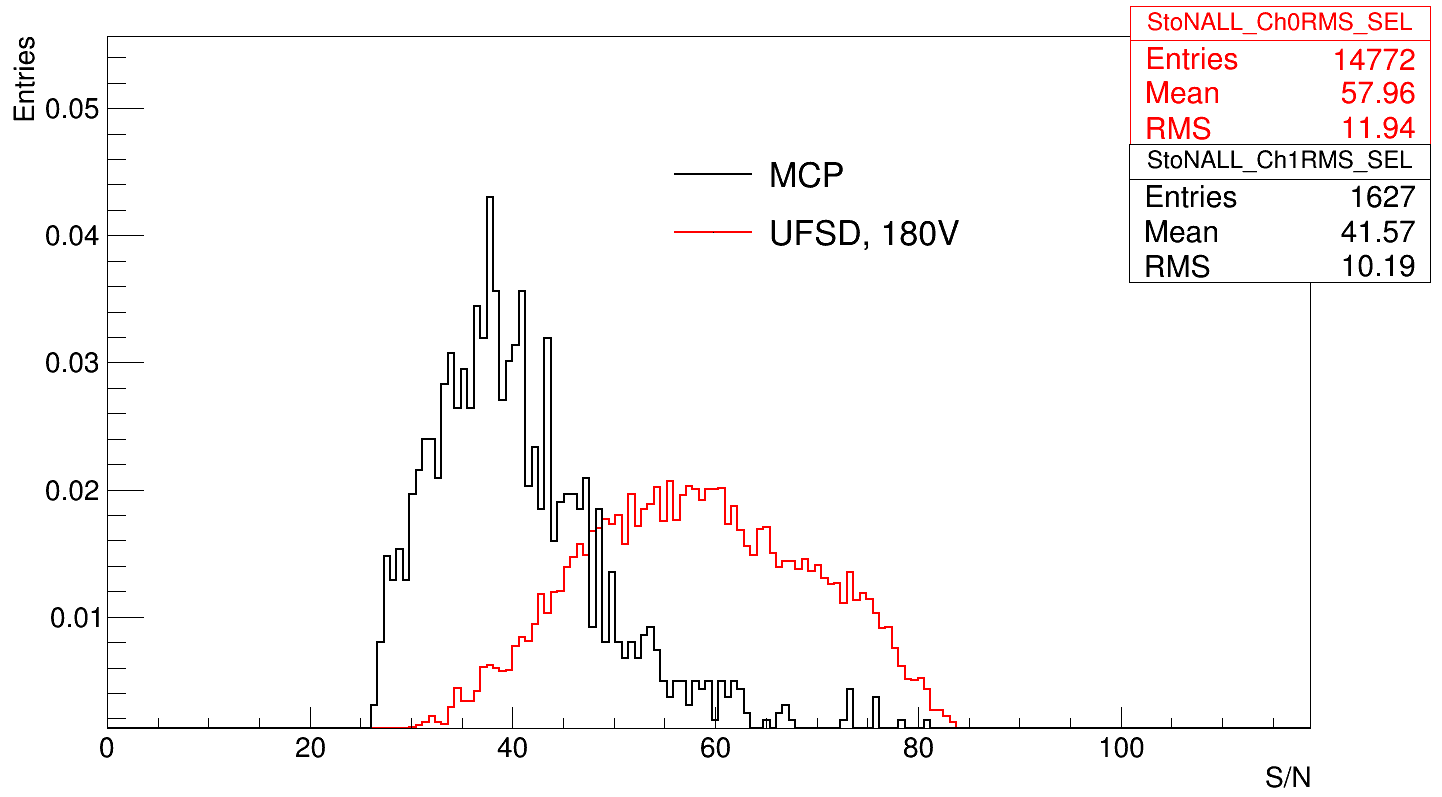}
\caption{Signal to Noise ratio of the MCP and of the 2\,mm$^2$ UFSD pixel. }
\label{ston}
\end{figure}
The particle rate was $\sim 10^3\,$/mm$^2$, the HV on the UFSD was set initially at 180\,V, which is the maximum voltage before pixels breakdown, and varied down to 140\,V.
The maximum current allowed in the present measurement was 0.1\,mA. 
A~screen shot from the oscilloscope with the signals from the MCP and the UFSD detectors is shown in Figure~\ref{edisp}.

\begin{figure}[!h]
\centering
\includegraphics[width=0.7\linewidth]{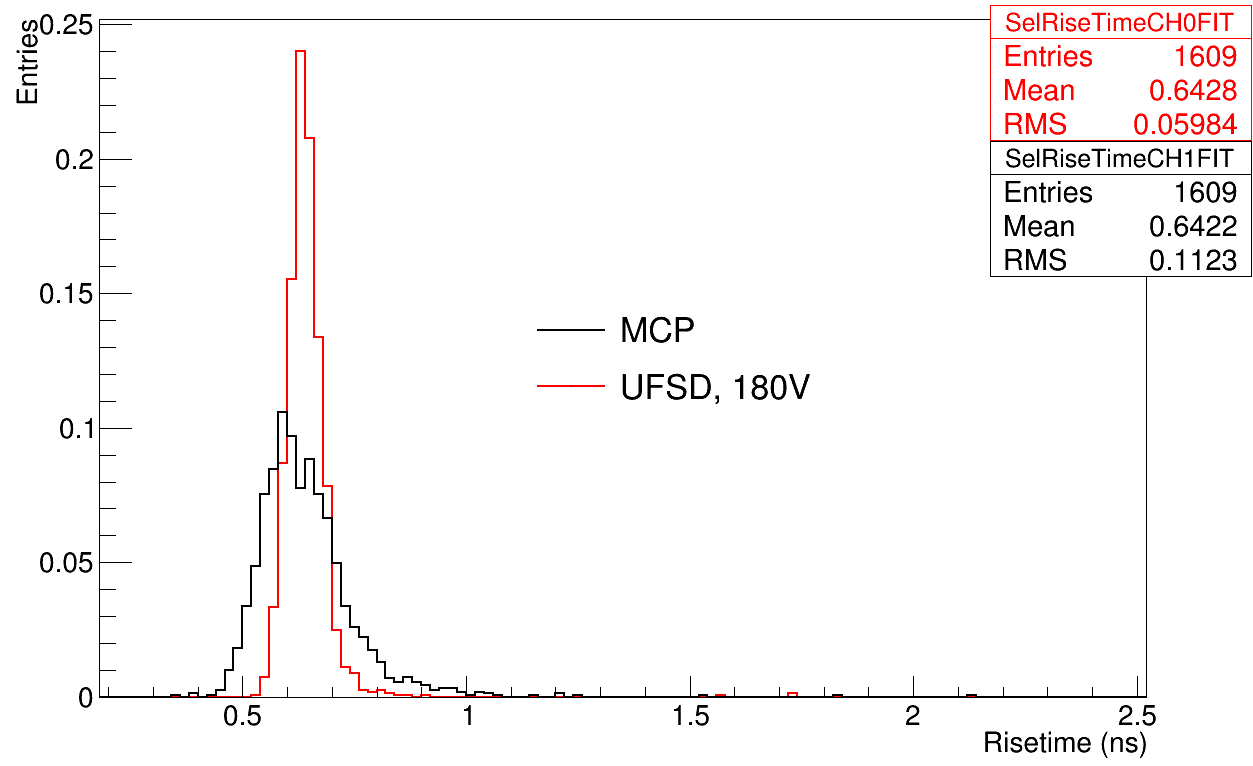}
\caption{Risetime of the MCP and of the 2\,mm$^2$ UFSD pixel.}
\label{rise}
\end{figure}

The UFSD pixels that we tested have an area ranging between 1.8\,mm$^2$ and 14\,mm$^2$. 
The $2.2\, {\rm mm}^2$ UFSD pixel shows an average S/N of $\sim $60 (Figure~\ref{ston}) and a risetime of 0.6\,ns (Figure~\ref{rise}) . 

The UFSD S/N curve for the events used in this analysis does not show the typical Landau curve tail ; this is due to the saturation of $\sim\,$10\% of the signals and may include the effect of a non linearity in the modified amplification chain.

\begin{figure}[!h]
\centering
\includegraphics[width=0.7\linewidth]{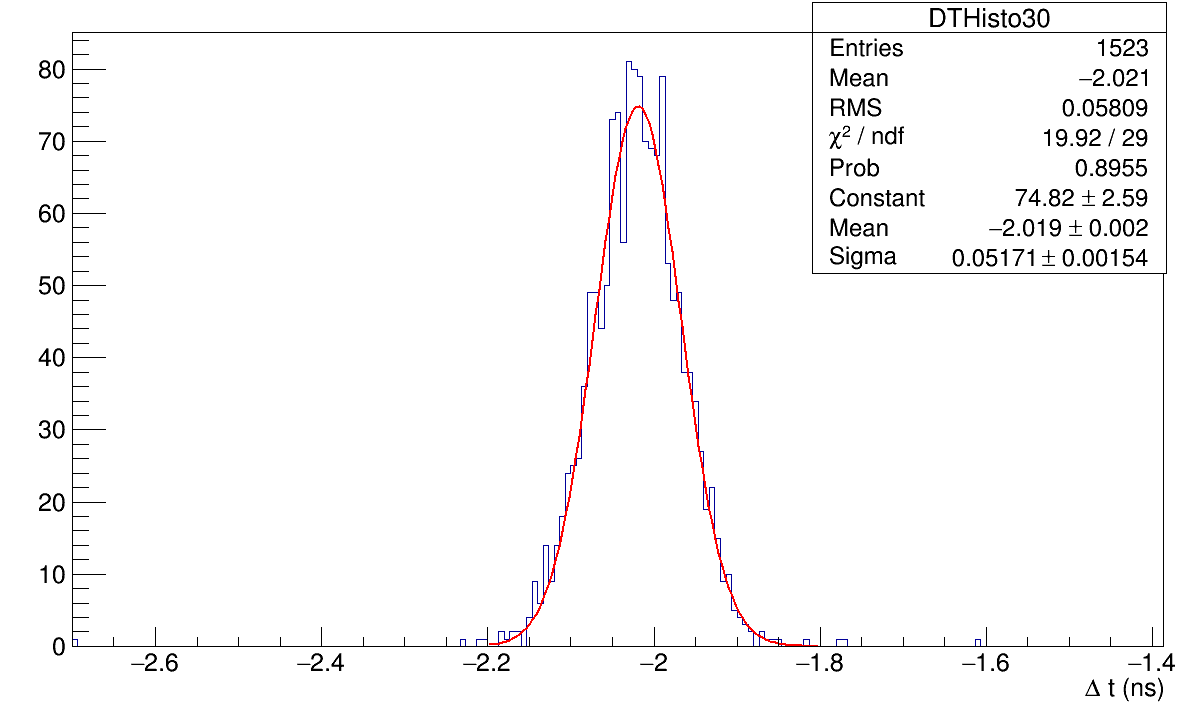}
\caption{Difference of the arrival time measured by the MCP and by the  2.2\,mm$^2$ UFSD pixel biased at 180\,V.}
\label{deltaT}
\end{figure}

Signals are recorded with a 20 Gsa/s DSO9254A  Agilent oscilloscope.
The time difference between the MCP and the $2.2\, {\rm mm}^2$ UFSD pixel is shown in Figure~\ref{deltaT}. The difference is computed off-line by using a constant fraction discrimination with a threshold at 30\% of the maximum for both the UFSD and the MCP signal. 

The MCP time precision was obtained from other measurements and is $(40\pm 5)\,$ps.

The results of the measurements  are summarized in Table~\ref{table:results}.

\begin{figure}[!h]
\centering
\includegraphics[width=0.7\linewidth]{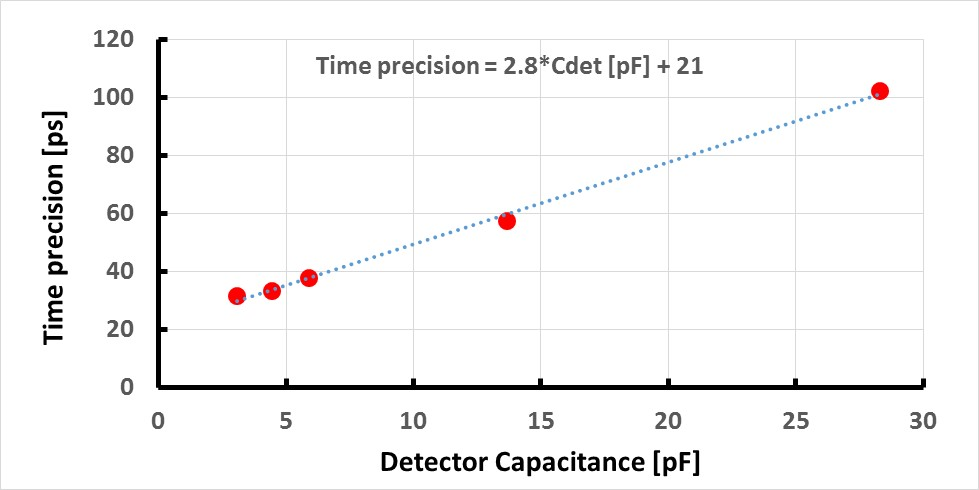}
\caption{UFSD time precision as a function of the pixel capacitance for a bias of 180V.}
\label{summary-C}
\end{figure}

\begin{figure}[!h]
\centering
\includegraphics[width=0.7\linewidth]{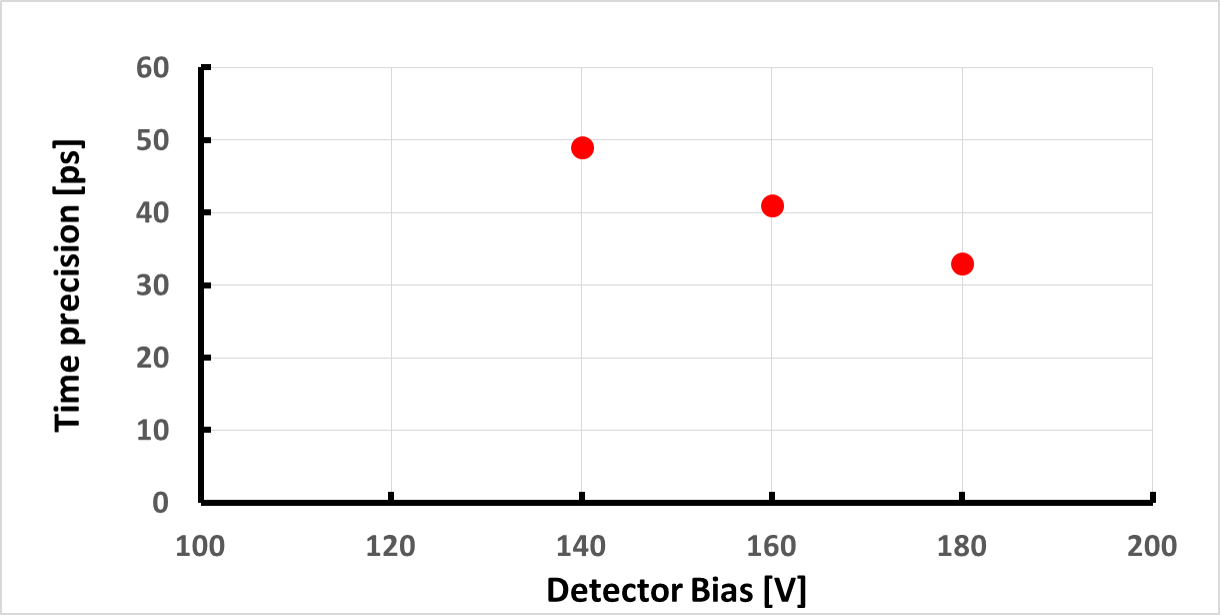}
\caption{UFSD time precision ( 2.2\,mm$^2$ pixel) as a function of the applied bias Voltage.}
\label{summaryhv}
\end{figure}

Figures~\ref{summary-C} and~\ref{summaryhv} show the UFSD  time precision as a function of the pixel capacitance and of the applied bias voltage respectively; the second set of measurements was performed on the pixel with an area of 2.2\,mm$^2$.
The precision of the measurement is mainly due to the uncertainty with which we know the MCP time precision.

\begin{table}[h]
\centering
\caption{Results of the time precision measurements as a function of the pixel capacitance, pixel surface area and of the applied bias Voltage. The uncertainty on the measured values is of $\sim 5\,$ps and depends essentially on the uncertainty of the MCP reference measurement.}
\begin{tabular}{c c c c}
\hline
  Surface  & Capacitance & HV & Time precision\\
  {[mm$^2$]} & [pF] & [V] & [ps]\\
 \hline

 1.8 & 3.1 & 180 & 32\\
 2.2 & 4.4 & 180 &  33\\
 3.0 & 6.0 & 180 & 38 \\
 7.0 & 14 & 180 & 57\\
 14 & 28  & 180 & 102\\
\hline
2.2 & 4.4 & 140 &  49\\
2.2 & 4.4 & 160 &  41\\
2.2 & 4.4 & 180 &  33\\
\hline
\end{tabular}

\label{table:results}
\end{table}

The trend of the measurements suggests  that a time precision of less than 30\,ps could be reached for the  smallest area pixel biased at 200\,V.

\section{Conclusions}
\label{sec:concl}
Here we described the timing performance of a 50\,$\mu$m thick UFSD detector on a beam of minimum ionizing particles. 
A time precision in the range of 30-100\,ps has been measured, depending on the pixel capacitance .
The UFSD technology will be used by TOTEM experiment in  the vertical RPs together with scCVD sensors.

\section*{Acknowledgments}
\label{sec:ack}
We thank Florentina Manolescu and Jan Mcgill for the realization of the unusual bonding of the sensors.
Support for some of us to travel to CERN for the beam tests was provided by AIDA-2020-CERN-TB-2016-11.
This work was supported by the institutions listed on the front page and also by the project LM2015058 from the Czech Ministry of Education Youth and Sports.
Part of this work has been financed by the European Union’s Horizon 2020 Research and Innovation funding program, under Grant Agreement no. 654168 (AIDA-2020) and Grant Agreement no. 669529 (ERC UFSD669529), and by the Italian Ministero degli Affari Esteri and INFN Gruppo I and V.
The design was supported by National program of sustainability LO1607 Rice-Netesis of the Ministry of Education, Youth and Sports, Czech Republic.

\bibliographystyle{unsrt}

\bibliography{Diam-2016-DD-base-J}


\end{document}